\documentclass[preprint,prd,showpacs]{revtex4}

\usepackage{amsfonts, amssymb, amsthm, amsmath}
\usepackage{epsfig}
\usepackage{epstopdf}
\usepackage{amsfonts, amssymb, amsthm}

\renewcommand{\theequation}{\arabic{section}.\arabic{equation}}

\newcommand{\be}{\begin{equation}}

\newcommand{\ee}{\end{equation}}

\newcommand{\bea}{\begin{eqnarray}}

\newcommand{\eea}{\end{eqnarray}}

\newcounter{orange}
\renewcommand{\theorange}{\alph{orange}}

\begin{document}
\title{ From the discrete Weyl -- Wigner formalism for symmetric ordering to a number -- phase Wigner function}

\author{Maciej Przanowski} \email{ maciej.przanowski@p.lodz.pl} 
\affiliation{  Institute of Physics,   \L\'{o}d\'{z} University of Technology, \\ W\'{o}lcza\'{n}ska 219, 90-924 \L\'{o}d\'{z}, Poland}

\author{Jaromir Tosiek} \email{jaromir.tosiek@p.lodz.pl}
\affiliation{  Institute of Physics,   \L\'{o}d\'{z} University of Technology, \\ W\'{o}lcza\'{n}ska 219, 90-924 \L\'{o}d\'{z}, Poland}

\date{\today}

\begin{abstract}
The general Weyl -- Wigner formalism in finite dimensional phase spaces is investigated. Then this formalism is specified to the case of symmetric ordering of operators in an odd -- dimensional Hilbert space. A respective Wigner function on the discrete phase space is found and the limit, when the dimension of Hilbert space tends to infinity, is considered. It is shown that this limit gives the number -- phase Wigner function in quantum optics. Analogous results for the `almost' symmetric ordering in an even -- dimensional Hilbert space are obtained. Relations between the discrete Wigner functions introduced in our paper and some other discrete Wigner functions appearing in literature are studied.

\end{abstract}


\maketitle

%%%%%%%%%%%%%%%%%%%%%%%%%%%%%%%%%%%%%%%%%%%%%%%%%%%%%%%%%%%%%%%%%%%%%%%%%%%%%%%%%%%%%%%%%%%%%%%%%%%%%%%%%%%%%%%%%%%%%%%%%%%%%%%%%%%%%%%%%%%%%%%%%%%%%%%%%%%%%%%%%%%%%%%%%%%%%%%%%%%%%%%%%%%%%%%%%%%%%%%%%%%%%%%%%%%%%%%%%%%%%%%%%%%%%%%%%%%%%%%%%%%%%%%%%%%%%%%%%%%%%%%%%%%%%%%%%%%%%%%%%%%%%%%%%%%%%%%%%%%%%%%%%%%%%%%%%%%%%%%%%%%%%%%%%%%%%%%%%%%%%%%%%%%%%%%%%%%%%%%%%%%%%%%%%%%%%%%%%%%%%%%%%%%%%%%%%%%%%%%%%%%%%%%%%%%%%%%%%%%%%%%%%%%%%%%%%%%%%%%%%%%%%%%%%%%%%%%%%%%%%%%%%%%%%%%%%%%%%%%%%%%%%%%%%%%%%%%%%%%%%%%%%%%%%%%%%%%%%%%%%%%%%%%%%%%%%%%%%%%%%%%%%%%%%%%%%%%%%%%%%%%%%%%%%%%%%%%%%%%%%%%%%%%%%%%%%%%%%%%%%%%%
\section{Introduction}
\label{sec1}
In the previous works \cite{PB13, PBT14, PTGT16, PB15} we have investigated some aspects of the concept of quantum phase. In particular we were able to define the number -- phase Wigner function \cite{PBT14, PB15} which in a natural way follows from the general Weyl -- Wigner formalism as specified to the symmetric ordering of operators. In the present paper we are going to develop the discrete Weyl -- Wigner formalism for a finite -- dimensional Hilbert space and then, in particular to find the respective discrete Wigner function for the symmetric ordering. Having done that one can easily obtain the number -- phase Wigner function by taking the limit when the dimension of Hilbert space tends to infinity. Quantum mechanics in finite dimensions has attracted a great deal of interest for many years \cite{W31, S60, ST76, S76, S77, SS78, JS81, ST84, F87, W87, CCSS88, VP90, L95, L96, V97, MLI01, THHH01, HTHH02, MPS02, CEMMMS05, CEMMMS06, HPB12} (see also the references in \cite{CEMMMS06, HPB12}).

One of  the main areas of investigation in this issue is the discrete phase -- space formalism i.e. the {\it discrete Weyl -- Wigner formalism} which consists in defining a {\it generalized Weyl application} and a {\it Wigner function}. The generalized Weyl application assigns uniquely an operator on the finite -- dimensional Hilbert space to any function given on the corresponding phase -- space grid. The Wigner function is a real function on this phase -- space grid representing a quantum state in the given finite -- dimensional Hilbert space. 

These two tasks are usually achieved simultaneously by the use of appropriate phase -- point operators, which are also called the {\it Stratonovich -- Weyl quantizer} or the {\it Fano operators} \cite{W87, CCSS88, VP90, THHH01, HTHH02, MPS02, CEMMMS06,  HPB12, PBT14}. We analyse this method in  detail in Sec. \ref{sec2}, so here we make only an important remark. Namely, it is obvious that both:   features of quantization and  properties of a Wigner function are determined by  assumed   phase -- point operators. Therefore one can decide to sacrifice a transparent rule of quantization (the operator ordering) for some desired properties of the Wigner function. This is just the approach proposed in \cite{W87, CCSS88, VP90, L95, L96, MLI01, HTHH02, MPS02}, where it is assumed that the Wigner function has the correct marginal distributions for all tilted lines on the phase -- space grid. The philosophy of our paper is slightly different. We are rather prone to sacrifice some properties of the Wigner function (e.g. the property mentioned above) in order to get a simple quantization rule (see also \cite{THHH01, CEMMMS06}) and  to get a transparent method for obtaining the number -- phase Wigner function when the dimension of the grid tends to infinity.

The paper is organized as follows. In Sec. \ref{sec2} we develop the general Weyl -- Wigner formalism in finite dimensions and present a general theory of the Stratonovich -- Weyl quantizer. In Sec. \ref{sec3} we deal with odd -- dimensional Hilbert space ${\cal H}^{(s+1)}, \; s+1=2N+1, \; N \in {\mathcal N} $ and we specify the Stratonovich -- Weyl quantizer to get the symmetric ordering of operators. Then we define the corresponding Wigner function and show that the limit $ N \rightarrow \infty$ leads to the number -- phase Wigner function given in \cite{PBT14, PB15}. The relation between our Wigner function and the one introduced in \cite{W87, CCSS88, VP90, L95, L96, HTHH02, MPS02} is studied in Sec. \ref{sec4}. The case of even -- dimensional Hilbert spaces ${\cal H}^{(s+1)}, \; s+1=2N, \; N \in {\mathcal N} $ is investigated in Sec. \ref{sec5}. We find the Stratonovich -- Weyl quantizer which defines the {\it almost symmetric ordering} of operators. Then relations between this Wigner function and that found by U. Leonhardt \cite{L95, L96} (see also \cite{MPS02}) are analysed. Some conclusions end the paper.

%%%%%%%%%%%%%%%%%%%%%%%%%%%%%%%%%%%%%%%%%%%%%%%%%%%%%%%%%
%%%%%%%%%%%%%%%%%%%%%%%%%%%%%%%%%%%%%%%%%%%%%%%%%%%%%%%%
\section{The general Weyl -- Wigner formalism in finite dimensions}
\label{sec2}

Consider an $(s+1)$ -- dimensional Hilbert space ${\cal H}^{(s+1)}$ equipped with an orthonormal basis 
$\big\{ \big|0\big>, \big|1\big>, \ldots,  \big|s\big> \big\}, \;\; \big<n \big|n' \big>= \delta_{nn'}, \;\; n,n'=0,1,\ldots,s.$ Inspired by the celebrated Pegg -- Barnett construction of the quantum phase \cite{PB88, PB89, BP89} we introduce another orthonormal basis $\big\{ \big|\phi_0\big>, \big|\phi_1\big>, \ldots,  \big|\phi_s\big> \big\},$
\be
\label{2.1}
\big|\phi_m\big>:= \frac{1}{\sqrt{s+1}} \sum_{n=0}^s \exp(i n \phi_m) \big|n\big>, 
\ee
\[ 
\big<\phi_m \big| \phi_{m'} \big>= \delta_{mm'}\;\;, \;\; m,m'=0,1,\ldots,s
\]
with
\be
\label{2.2}
\phi_m= \phi_0 + \frac{2 \pi}{s+1}m, \;\; m=0,1,\ldots,s.
\ee
Define then two hermitian operators
\be
\label{2.3}
\hat{n}:= \sum_{n=0}^s n \big| n \big> \big< n \big|
\ee
and
\be
\label{2.4}
\hat{\phi}:= \sum_{m=0}^s \phi_m \big| \phi_m \big> \big< \phi_m \big|
\ee
which enable us to construct the following unitary operators
\be
\label{2.5}
\hat{V}:= \exp \left(i \frac{2 \pi}{s+1} \hat{n} \right)
\ee
satisfying $\hat{V}^{s+1}= \hat{\bf 1}$
and
\be
\label{2.6}
\hat{U}:= \exp (i \hat{\phi} )
\ee
fulfilling the equality $\hat{U}^{s+1}= \exp \Big\{ i (s+1) \phi_0 \Big\} \hat{\bf 1}. $

As is well known, the analogous operators play the fundamental role in the discrete Weyl -- Wigner formalism \cite{W31, S60, ST76, S76, S77, SS78, JS81, ST84, THHH01, HTHH02, MPS02, HPB12}.

One quickly finds that the operator $\hat{U}$ can be rewritten in the form
\be
\label{2.7}
\hat{U}= \sum_{n=0}^{s-1} \big|n \big> \big< n+1\big| + \exp \Big\{ i (s+1) \phi_0\Big\} \big|s \big> \big< 0 \big|.
\ee
Employing (\ref{2.7}) one gets Weyl's commutation relation
\be
\label{2.8}
\exp \left(-i \frac{2 \pi k l}{s+1} \right) \hat{U}^k \hat{V}^l= \hat{V}^l \hat{U}^k, \;\; k,l \in {\mathbb Z}
\ee
or
\be
\label{2.9}
\exp \left(-i \frac{ \pi k l}{s+1} \right) \hat{U}^k \hat{V}^l= \exp \left(i \frac{ \pi k l}{s+1} \right) \hat{V}^l \hat{U}^k, \;\; k,l \in {\mathbb Z}.
\ee
The main point is that the relation (\ref{2.9}) resembles the well known formula in $L^2({\mathbb R}^1)$
\[
\exp \big\{i (\lambda \hat{p} + \mu \hat{q}) \big\} = \exp \left(- \frac{i \hbar \lambda \mu}{2} \right) \exp (i \lambda \hat{p}) \exp (i \mu \hat{q})
\]
\be
\label{2.10}
 = \exp \left( \frac{i \hbar \lambda \mu}{2} \right)  \exp (i \mu \hat{q}) \exp (i \lambda \hat{p}), \;\; \lambda, \mu \in {\mathbb R},
\ee
with the commutation relation $[\hat{q}, \hat{p}]= i \hbar \hat{\bf 1}.$ Thus we expect that the operator $\exp \left(-i \frac{ \pi k l}{s+1} \right) \hat{U}^k \hat{V}^l$ will play in our further construction a role analogous to the unitary displacement operator $\exp \big\{i (\lambda \hat{p} + \mu \hat{q}) \big\} $ in the Weyl -- Wigner formalism in the space $L^2({\mathbb R}^1).$

Denote
\be
\label{2.11}
\hat{D}(k,l):=\exp \left(-i \frac{ \pi k l}{s+1} \right) \hat{U}^k \hat{V}^l, \;\; k,l \in {\mathbb Z}.
\ee
One easily shows that the operators $\hat{D}(k,l)$ satisfy the relations
 \setcounter{orange}{1}
\renewcommand{\theequation} {\arabic{section}.\arabic{equation}\theorange}
\be
\label{2.12a}
\hat{D}^+(k,l)=\hat{D}^{-1}(k,l)=\hat{D}(-k,-l), \;\;  k,l \in {\mathbb Z},
\ee
\addtocounter{orange}{1}
\addtocounter{equation}{-1}
\be
\label{2.12b}
{\rm Tr} \{\hat{D}(k,l)\}= (s+1) \delta_{k0} \delta_{l0}, \;\; 0 \leq k,l \leq s,
 \ee
 \addtocounter{orange}{1}
\addtocounter{equation}{-1}
\be
\label{2.12c}
{\rm Tr} \{\hat{D}(k,l)\hat{D}^+(k',l')\}= (s+1) \delta_{kk'} \delta_{ll'}, \;\; 0 \leq k,l,k',l' \leq s.
 \ee
 \renewcommand{\theequation} {\arabic{section}.\arabic{equation}}
 Define
 \be
 \label{2.13}
 \big|\phi_r\big>:= \frac{1}{\sqrt{s+1}} \sum_{n=0}^s \exp \left\{in \left(\phi_0 + \frac{2 \pi r}{s+1}  \right) \right\}
 \ee
 for any $r \in {\mathbb Z}.$ Then $\hat{D}(k,l)$ can be represented in the form
 \be
 \label{2.14}
 \hat{D}(k,l)=\exp \left(i \frac{ \pi k l}{s+1} \right) \sum_{m=0}^s \exp (i k \phi_m) \big|\phi_{m+l} \big> \big<\phi_m \big|, \;\; 0 \leq k,l \leq s.
 \ee
 The set of pairs $\big\{(\phi_m,n) \big\}_{m,n=0}^s$ constitutes the grid, which we consider as the discrete phase -- space denoted by $\Gamma^{(s+1)}$ and  associated with the Hilbert space ${\cal H}^{(s+1)}.$ Any function $f=f(\phi_m,n)$ on $\Gamma^{(s+1)}$ can be expanded in the Fourier sum
 \be
 \label{2.15}
 f(\phi_m,n)= \frac{1}{s+1} \sum_{k,l=0}^s \tilde{f}(k,l) \exp \left\{ i \left( k \phi_m + \frac{2 \pi}{s+1}ln\right)\right\}
 \ee
with
\be
\label{2.16}
\tilde{f}(k,l):= \frac{1}{s+1} \sum_{m,n=0}^s  f(\phi_m,n) \exp \left\{- i \left( k \phi_m + \frac{2 \pi}{s+1}ln\right)\right\}.
\ee
Following the (generalized) Weyl -- Wigner formalism in ${\mathbb R}^{2n}$ \cite{W31,W32,M64,C66,S57,F57,T83,TP95, PPT96} we define the {\it phase -- point operators} (the {\it Stratonovich -- Weyl quantizer}) on $\Gamma^{(s+1)}$ as
\be
\label{2.17}
\hat{\Omega}[{\cal K}](\phi_m, n):= \frac{1}{s+1} \sum_{k,l=0}^s {\cal K}(k,l)  \hat{D}(k,l),
\exp \left\{- i \left( k \phi_m + \frac{2 \pi}{s+1}l n\right)\right\}
\ee
where ${\cal K}(k,l)$ is called the {\it kernel} and it should have some properties, which are determined by expected properties of $\hat{\Omega}[{\cal K}](\phi_m, n).$

According to the Weyl quantisation rule we assign the operator $\hat{f}$ to a function $f=f(\phi_m,n)$ by the formula
\be
\label{2.18}
\hat{f}= \frac{1}{s+1}\sum_{k,l=0}^s \tilde{f}(k,l) \hat{D}(k,l)= 
 \frac{1}{s+1}\sum_{m, n=0}^s f(\phi_m,n) \hat{\Omega}[{\cal K}](\phi_m,n),
\ee
where $\tilde{f}(k,l)$ is given by (\ref{2.16}). Employing (\ref{2.12c}) one gets from (\ref{2.18})
\be
\label{2.19}
{\cal K}[k,l]\tilde{f}(k,l)= {\rm Tr} \left\{ \hat{f} \hat{D}^+(k,l)\right\}.
\ee
Therefore we conclude that Eq. (\ref{2.18}) gives a one -- to -- one  correspondence between functions on $\Gamma^{(s+1)}$ and operators on ${\cal H}^{(s+1)}$ if and only if
\be
\label{2.20}
{\cal K}(k,l) \neq 0 \;\;\; \forall\; 0 \leq k,l \leq s.
\ee
From now on we assume that (\ref{2.20}) holds true. One wants the operator $\hat{f}$ assigned to $f=f(\phi_m,n)$ by (\ref{2.18}) to be hermitian for any real function $f(\phi_m,n).$ This is true iff $\hat{\Omega}[{\cal K}]$ is hermitian
\be
\label{2.21}
\hat{\Omega}[{\cal K}](\phi_m,n)=  \hat{\Omega}^+[{\cal K}](\phi_m,n), \;\; \forall \,(\phi_m,n) \in \Gamma^{(s+1)}.
\ee
From (\ref{2.17}), using also (\ref{2.12a}) one infers that the condition (\ref{2.21}) is fulfilled if and only if the kernel ${\cal K}$ fulfills the following conditions
 \setcounter{orange}{1}
\renewcommand{\theequation} {\arabic{section}.\arabic{equation}\theorange}
\be
\label{2.22a}
{\cal K}^*(k,l)= (-1)^{s+1+k+l}{\cal K}(s+1-k,s+1-l),\;\;\;  1 \leq k,l \leq s,
\ee
\addtocounter{orange}{1}
\addtocounter{equation}{-1}
\be
\label{2.22b}
{\cal K}^*(0,l)={\cal K}(0,s+1-l), \;\;\;  1 \leq l \leq s,
\ee
\addtocounter{orange}{1}
\addtocounter{equation}{-1}
\be
\label{2.22c}
{\cal K}^*(k,0)= {\cal K}(s+1-k,0),\;\;\;  1 \leq k \leq s,
\ee
\addtocounter{orange}{1}
\addtocounter{equation}{-1}
\be
\label{2.22d}
{\cal K}^*(0,0)={\cal K}(0,0),
\ee
\renewcommand{\theequation} {\arabic{section}.\arabic{equation}}
where the asterix `$^{*}$' stands for the complex conjugation.

The further natural assumption is that if $f=f(\phi_m,n)= 1,$ then the respective operator should be $\hat{f}=\hat{\bf 1}.$  From (\ref{2.18}) with (\ref{2.17}) one has
\[
\hat{\bf 1}= \frac{1}{s+1}\sum_{m,n=0}^s 1 \cdot \hat{\Omega}[{\cal K}](\phi_m,n)
\]
\[
=\frac{1}{(s+1)^2}\sum_{k,l=0}^s \left(\sum_{m,n=0}^s \exp \left\{ -i \left( k \phi_m + \frac{2 \pi}{s+1}ln \right)  \right\} \right)
{\cal K}(k,l) \hat{D}(k,l)
\]
\be
\label{2.23}
=\sum_{k,l=0}^s \delta_{k0} \delta_{l0} {\cal K}(k,l) \hat{D}(k,l) = {\cal K}(0,0) \cdot \hat{\bf 1}.
\ee
Hence
\be
\label{2.24}
{\cal K}(0,0)=1.
\ee
We assume also that if the function $f$ on $\Gamma^{(s+1)}$ depends on $\phi_m$ only, $f=f(\phi_m),$ then the respective operator $\hat{f}=f(\hat{\phi}).$ Analogously, if  the function $f$ depends exclusively on $n,$ $f=f(n),$ then the respective operator is $\hat{f}=f(\hat{n}).$

Consider any $f=f(\phi_m).$ Taking $\hat{f}=f(\hat{\phi})$ one has
\[
\hat{f}=f(\hat{\phi}) = \frac{1}{s+1}\sum_{m,n=0}^s  f(\phi_m) \hat{\Omega}[{\cal K}](\phi_m,n)
\]
\[
=\frac{1}{(s+1)^2}\sum_{k,l=0}^s  \left[ \left( \sum_{m=0}^s  f(\phi_m) \exp \left( -ik \phi_m\right) \right)
\left(\sum_{n=0}^s  \exp \left( -i \frac{2 \pi}{s+1} ln\right)\right)
{\cal K}(k,l) \hat{D}(k,l)
  \right]
\]
\[
=\frac{1}{s+1}\sum_{m,k=0}^s {\cal K}(k,0)  f(\phi_m) \exp \left( -ik \phi_m\right) \hat{D}(k,0).
\]
Using (\ref{2.14}) we get
\[
\hat{f}=f(\hat{\phi}) = \sum_{m=0}^s f(\phi_m) \big|\phi_m  \big> \big<\phi_m \big| = 
\frac{1}{s+1}\sum_{m,k,r=0}^s {\cal K}(k,0)  f(\phi_m) \exp \left\{ik (\phi_r - \phi_m) \right\}
\big|\phi_r  \big> \big<\phi_r \big|. 
\]
Hence
\[
f(\phi_r) = \frac{1}{s+1} \sum_{k=0}^s \left[ \left(  \sum_{m=0}^s f(\phi_m)  \exp \left( -ik \phi_m\right)\right) 
{\cal K}(k,0) \exp (ik \phi_r )
\right]
\]
for any function $f=f(\phi_m).$ This is true iff
\be
\label{2.25}
{\cal K}(k,0)=1 \;\;\; \forall \; 0  \leq k \leq s.
\ee
Analogously, assuming that for any $f=f(n)$ the respective operator is $\hat{f}=f(\hat{n})$ we find
\be
\label{2.26}
{\cal K}(0,l)=1 \;\;\; \forall \; 0  \leq l \leq s.
\ee
Observe that the conditions (\ref{2.25}) and (\ref{2.26}) yield (\ref{2.24}) and also (\ref{2.22b}), (\ref{2.22c}) and (\ref{2.22d}). Gathering, {\it we will assume that the kernel ${\cal K}$ satisfies the conditions (\ref{2.20}), (\ref{2.22a}),  (\ref{2.25}) and (\ref{2.26}).} 

So $\hat{\Omega}[{\cal K}]$ is hermitian. Moreover from (\ref{2.17}) and (\ref{2.12b}) one quickly gets
\be
\label{2.27}
{\rm Tr} \left\{\hat{\Omega}[{\cal K}] (\phi_m,n)\right\}= {\cal K}(0,0)=1.
\ee

The following properties of $\hat{\Omega}[{\cal K}]$ can be also easily proved:
\setcounter{orange}{1}
\renewcommand{\theequation} {\arabic{section}.\arabic{equation}\theorange}
\be
\label{2.28a}
\frac{1}{s+1} \, \sum_{n=0}^s \hat{\Omega}[{\cal K}] (\phi_m,n)= \big| \phi_m \big> \big< \phi_m\big|,
\ee
\addtocounter{orange}{1}
\addtocounter{equation}{-1}
\be
\label{2.28b}
\frac{1}{s+1}\, \sum_{m=0}^s \hat{\Omega}[{\cal K}] (\phi_m,n)= \big| n \big> \big< n \big|,
\ee
\addtocounter{orange}{1}
\addtocounter{equation}{-1}
\be
\label{2.28c}
\frac{1}{s+1} \, \sum_{m,n=0}^s \hat{\Omega}[{\cal K}] (\phi_m,n)= \hat{\bf 1}.
\ee
\renewcommand{\theequation} {\arabic{section}.\arabic{equation}}
Finally, from (\ref{2.17}) with (\ref{2.21})  and (\ref{2.12c})   we have
\[
{\rm Tr} \left\{\hat{\Omega}[{\cal K}] (\phi_m,n)
\hat{\Omega}[{\cal K}] (\phi_{m'},n')
\right\}
\]
\be
\label{2.29}
= \frac{1}{s+1} \, \sum_{k,l=0}^s |{\cal K}(k,l)|^2 \exp \left\{-i \left[ k(\phi_m -\phi_{m'})  + \frac{2 \pi}{s+1} l(n-n')\right] \right\}.
\ee
In particular, the general formula (\ref{2.29}) reduces to the form
\be
\label{2.30}
{\rm Tr} \left\{\hat{\Omega}[{\cal K}] (\phi_m,n)
\hat{\Omega}[{\cal K}] (\phi_{m'},n')
\right\}= (s+1) \delta_{m\, m'} \delta_{n \, n'}
\ee
if and only if
\be
\label{2.31}
|{\cal K}(k,l)|=1 \;\;\; \forall \; 0 \leq k,l \leq s.
\ee
In the present paper we do not assume, in general, that (\ref{2.30}) or, equivalently, (\ref{2.31}) hold true.

Inserting (\ref{2.19}) into (\ref{2.15}) one gets an inverse  formula to (\ref{2.18})
\be
\label{2.32}
f(\phi_m, n) = \frac{1}{s+1} \sum_{k,l=0}^s \frac{1}{{\cal K}(k,l)} \exp \left\{ i \left(k \phi_m + \frac{2 \pi}{s+1} l n  \right)\right\} \cdot {\rm Tr} \left\{ \hat{f}\, \hat{D}^+ (k,l)\right\}.
\ee
From (\ref{2.17}) we easily find $\hat{D}(k,l)$ as
\be
\label{2.33}
\hat{D}(k,l)= \frac{1}{s+1} \frac{1}{{\cal K}(k,l)} \sum_{m,n=0}^s
 \exp \left\{ i \left(k \phi_m + \frac{2 \pi}{s+1} l n  \right)\right\}
 \hat{\Omega}[{\cal K}] (\phi_m,n).
\ee
Substituting (\ref{2.33}) into (\ref{2.32}) one gets
\[
f(\phi_m, n) = \frac{1}{(s+1)^2}  \sum_{k,l,m',n'=0}^s \frac{1}{|{\cal K}(k,l)|^2}
\exp \left\{i \left[ k(\phi_m -\phi_{m'})  + \frac{2 \pi}{s+1} l(n-n')\right] \right\}
\]
\be
\label{2.34}
 \times
{\rm Tr} \left\{ \hat{f}\,\hat{\Omega}[{\cal K}] (\phi_{m'},n')
\right\}.
\ee
In the case when the condition (\ref{2.31}) (or, equivalently (\ref{2.30})) is satisfied, Eq. (\ref{2.34}) reduces to a simple formula
\be
\label{2.35}
f(\phi_m, n) = {\rm Tr} \left\{ \hat{f}\,\hat{\Omega}[{\cal K}] (\phi_{m},n)
\right\}.
\ee
Now we are at the position where a quasiprobability distribution relevant to the quantisation rule (\ref{2.18}) can be introduced. According to this rule the average value of the observable represented by $\hat{f}$ in a state defined by the density operator $\hat{\varrho}$ reads
\be
\label{2.36}
\big< \hat{f}\big> = {\rm Tr} \left\{ \hat{f} \, \hat{\varrho} \right\} = 
\sum_{m,n=0}^s f(\phi_m, n) \frac{1}{s+1}  {\rm Tr} \left\{ \hat{\varrho}\,\hat{\Omega}[{\cal K}] (\phi_{m},n) \right\}.
\ee
The last formula suggests a natural definition of the respective quasiprobability distribution as
\be
\label{2.37}
\varrho_W [{\cal K}] (\phi_{m},n):= \frac{1}{s+1}  \left\{ \hat{\varrho}\,\hat{\Omega}[{\cal K}] (\phi_{m},n) \right\}.
\ee
Thus (\ref{2.36}) can be rewritten in the form
\be
\label{2.38}
\big< \hat{f}\big> = \sum_{m,n=0}^s f(\phi_m, n) \varrho_W [{\cal K}] (\phi_{m},n).
\ee
The function $ \varrho_W [{\cal K}] (\phi_{m},n)$ will be called the {\it Wigner function of the state $\hat{\varrho}$  for the kernel ${\cal K}$}.

First, since $\hat{\varrho}^+=\hat{\varrho}$ and $\hat{\Omega}^+[{\cal K}]=\hat{\Omega}[{\cal K}],$ the respective Wigner function $\varrho_W [{\cal K}]$ is real
\be
\label{2.39}
\varrho_W^* [{\cal K}](\phi_{m},n)=\varrho_W [{\cal K}](\phi_{m},n)\;\;\; \forall \, (\phi_{m},n) \in \Gamma^{(s+1)}.
\ee
Then by (\ref{2.28c}) we have
\be
\label{2.40}
\sum_{m,n=0}^s \varrho_W [{\cal K}](\phi_{m},n) = {\rm Tr} \left\{\hat{\varrho} \right\}=1.
\ee
Moreover, from (\ref{2.28a}) one gets the marginal distribution
\be
\label{2.41}
\sum_{n=0}^s \varrho_W [{\cal K}](\phi_{m},n)= {\rm Tr} \left\{\hat{\varrho}  \big|\phi_m\big> \big<\phi_m \big| \right\}=
\big<\phi_m \big| \hat{\varrho} \big|\phi_m\big>
\ee
and from  (\ref{2.28b})  another marginal distribution
\be
\label{2.42}
\sum_{m=0}^s \varrho_W [{\cal K}](\phi_{m},n)= {\rm Tr} \left\{\hat{\varrho}  \big|n\big> \big<n \big| \right\}=
\big<n \big| \hat{\varrho} \big|n\big>.
\ee
We are going to find a formula inverse to (\ref{2.37}) i.e. we want to reveal the state $\hat{\varrho}$ from a given Wigner function $\varrho_W [{\cal K}].$ To this end we consider an operator $ \big|\phi_r \big> \big<\phi_{r'} \big|$, $  0 \leq r, r' \leq s.$ 
Let $f_{r r'}(\phi_{m},n)$ be a function corresponding to this operator according to (\ref{2.32}). Employing (\ref{2.14}) one easily obtains
\[
f_{r r'}(\phi_{m},n)= \frac{1}{s+1} \sum_{k,l=0}^s \frac{1}{{\cal K}(k,l)} \exp \left\{i \left(k \phi_m +\frac{2 \pi}{s+1}ln \right) \right\} \big<\phi_{r'} \big|\hat{D}^+(k,l) \big|\phi_r\big>
\]
\be
\label{2.43}
= \frac{1}{s+1} \exp \left\{i \frac{2 \pi}{s+1}(r-r') n \right\} \sum_{k=0}^s \frac{1}{{\cal K}(k,r-r')}
 \exp \left\{i \frac{2 \pi k}{s+1} \left( m- \frac{r+r'}{2} \right)  \right \}, \;\; {\rm for} \;\; r \geq r' .
\ee
Then by (\ref{2.37}), (\ref{2.38}) and (\ref{2.43}) we have
\[
\Big<  \big|\phi_r\big> \big<\phi_{r'} \big| \Big>=  {\rm Tr} \left\{\hat{\varrho} \big |\phi_r\big> \big<\phi_{r'} \big| \right\}=
\big<\phi_{r'} \big| \hat{\varrho}\big|\phi_r\big> = 
\sum_{m,n=0}^s f_{r r'}(\phi_{m},n) \varrho_W [{\cal K}](\phi_{m},n)
\]
\[
=  \frac{1}{s+1} \sum_{k,m,n=0}^s \frac{1}{{\cal K}(k,r-r')}
 \exp \left\{i \frac{2 \pi k}{s+1} \left( m- \frac{r+r'}{2} \right)  \right \}
 \]
\be
\label{2.44}
\times
  \exp \left\{i \frac{2 \pi}{s+1}(r-r') n \right\}
  \varrho_W [{\cal K}](\phi_{m},n), \; \;\; r \geq r' .
\ee
Since $\big<\phi_{r'} \big| \hat{\varrho} \big|\phi_r\big>^* = \big<\phi_{r} \big| \hat{\varrho} \big|\phi_{r'}\big>,$ one can easily obtain the matrix elements $\big<\phi_{r'} \big| \hat{\varrho} \big|\phi_r\big>, \;\; r \geq r' $ from (\ref{2.44}) as
\[
\big<\phi_{r'} \big| \hat{\varrho} \big|\phi_r\big> = \big<\phi_{r} \big| \hat{\varrho} \big|\phi_{r'}\big>^*=
\frac{1}{s+1} \sum_{k,m,n=0}^s \frac{1}{{\cal K}^*(k,r'-r)}
 \exp \left\{-i \frac{2 \pi k}{s+1} \left( m- \frac{r+r'}{2} \right)  \right \}
\]
\be
\label{2.45}
\times
  \exp \left\{i \frac{2 \pi}{s+1}(r-r') n \right\}
  \varrho_W [{\cal K}](\phi_{m},n), \; \;\; r \leq r' .
\ee
The formulas (\ref{2.44}) and (\ref{2.45}) give all the matrix elements $\big<\phi_{r'} \big| \hat{\varrho} \big|\phi_r\big>$, $ 0 \leq r, r' \leq s$ and, consequently, define the state $\hat{\varrho}$ by the corresponding Wigner function $\varrho_W [{\cal K}].$

In terms of the phase -- space operator $\hat{\Omega}[{\cal K}]$ the matrix element $\big<\phi_{r'} \big| \hat{\varrho} \big|\phi_r\big>$ for any  $ 0 \leq r, r' \leq s$ takes the form (see (\ref{2.43}) with (\ref{2.33}))
\[
\big<\phi_{r'} \big| \hat{\varrho} \big|\phi_r\big> = \frac{1}{(s+1)^2}
\sum_{k,l,m,n,m',n'=0}^s \frac{1}{|{\cal K}(k,l)|^2}
\exp \left\{i \left[ k(\phi_m-\phi_{m'}) + \frac{2 \pi}{s+1}l(n - n')\right] \right\}
\]
\be
\label{2.46}
\times \big<\phi_{r'} \big| \hat{\Omega}[{\cal K}](\phi_{m'},n') \big|\phi_r\big>\, \varrho_W [{\cal K}](\phi_{m},n), 
\;\; 0 \leq r, r' \leq s.
\ee
Hence
\[
\hat{\varrho}= 
 \frac{1}{(s+1)^2}
\sum_{k,l,m,n,m',n'=0}^s \frac{1}{|{\cal K}(k,l)|^2}
\exp \left\{i \left[ k(\phi_m-\phi_{m'}) + \frac{2 \pi}{s+1}l(n - n')\right] \right\}
\]
\be
\label{2.47}
\times  \hat{\Omega}[{\cal K}](\phi_{m'},n') \, \varrho_W [{\cal K}](\phi_{m},n).
\ee
In particular, if the condition (\ref{2.31}) (or, equivalently, (\ref{2.30})) is satisfied then Eq. (\ref{2.47}) reduces to a simple form
\be
\label{2.48}
\hat{\varrho}= \sum_{m,n=0}^s  \hat{\Omega}[{\cal K}](\phi_{m},n) \, \varrho_W [{\cal K}](\phi_{m},n).
\ee
In the next section we discuss a simple example of the kernel ${\cal K}(k,l)$ for an odd -- dimensional Hilbert space ${\cal H}^{(s+1)}.$ This kernel leads to the symmetric ordering of operators in (\ref{2.18}) i.e. for any function $f(\phi_m,n)= f_1(\phi_m) f_2(n)$ the assigned operator reads
\[
f_1(\phi_m) f_2(n) \longrightarrow \frac{1}{2} \left(f_1(\hat{\phi}) f_2(\hat{n}) +  f_2(\hat{n}) f_1(\hat{\phi}) \right).
\]
Moreover, it enables us to reach in a natural manner the limit $s \rightarrow \infty$ and, consequently, to find in this way the number -- phase Wigner function in quantum optics investigated in our previous works \cite{PBT14, PB15}.
%%%%%%%%%%%%%%%%%%%%%%%%%%%%%%%%%%%%%%%%%%%%%%%%%%%%%%%%
%%%%%%%%%%%%%%%%%%%%%%%%%%%%%%%%%%%%%%%%%%%%%%%%%%%%%%%%
\section{Kernel ${\cal K}(k,l)$ for an odd -- dimensional Hilbert space ${\cal H}^{(s+1)}$ in the symmetric ordering}
\label{sec3}
\setcounter{equation}{0}

We assume that $s=2N.$ Thus
\be
\label{3.1}
\dim {\cal H}^{(s+1)}= 2N+1, \;\; N \in {\mathcal N}.
\ee
One quickly shows that the function
\be
\label{3.2}
{\cal K}(k,l)= \cos \left( \frac{\pi kl}{2N+1}\right), \;\; 0 \leq k,l \leq 2N
\ee
fulfills all the conditions (\ref{2.20}), (\ref{2.22a}), (\ref{2.25}) and (\ref{2.26}). Therefore, the function (\ref{3.2}) can be taken as the kernel ${\cal K}(k,l)$ in the Weyl -- Wigner formalism considered in Sec. \ref{sec2}. In the present section we choose the kernel in the form (\ref{3.2}) and we will write simply  $\hat{\Omega}$ or $\varrho_W$ instead of 
$\hat{\Omega}[{\cal K}]$ or $\varrho_W[{\cal K}].$

Substituting (\ref{3.1}) and (\ref{3.2}) into (\ref{2.17}), then using  (\ref{2.14}) after some simple algebra manipulations one arrives at the result
\be
\label{3.3}
\hat{\Omega}(\phi_m,n)= \frac{2N+1}{2} \Big( \big|\phi_m\big> \big<\phi_m \big| n\big> \big<n \big|+ \big|n\big> \big<n \big|\phi_m\big> \big<\phi_m \big|\Big).
\ee
Such phase -- point operators lead to the symmetric ordering of operators i.e. for any function
$f(\phi_m,n)$ of the form
  $f(\phi_m,n)= f_1(\phi_m) f_2(n)$ the corresponding operator defined by (\ref{2.18}) reads
  \[
  \hat{f}= \frac{1}{2N+1} \sum_{m,n=0}^{2N}f_1(\phi_m) f_2(n) \frac{2N+1}{2}
  \Big( \big|\phi_m\big> \big<\phi_m \big| n\big> \big<n \big|+ \big|n\big> \big<n \big|\phi_m\big> \big<\phi_m \big|\Big)
  \]
  \be
  \label{3.4}
  = \frac{1}{2} \left(f_1(\hat{\phi}) f_2(\hat{n}) +  f_2(\hat{n}) f_1(\hat{\phi}) \right).
  \ee
  The Wigner function for the kernel (\ref{3.2}) can be easily found from the definition (\ref{2.37}) with (\ref{3.3}) One has
  \[
  \varrho_W(\phi_m,n)= \frac{1}{2N+1}\, {\rm Tr} \left\{\hat{\varrho } \,
  \frac{2N+1}{2}
  \Big( \big|\phi_m\big> \big<\phi_m \big| n\big> \big<n \big|+ |n\big> \big<n \big|\phi_m\big> \big<\phi_m \big|\Big)
  \right\}
  \]
\be
\label{3.5}
=\Re \Big\{ \big<n \big| \hat{\varrho} \big|\phi_m \big> \big< \phi_m \big| n \big>\Big\}.
\ee  
Inserting
\[
{\cal K}(k,r-r')= \cos  \left( \frac{\pi k(r-r')}{2N+1}\right) = \frac{1}{2}
\left[  \exp \left\{i \frac{\pi k(r-r')}{2N+1}\right\}   + \exp \left\{-i \frac{\pi k(r-r')}{2N+1}\right\}    \right]
\]
and performing some straightforward manipulations we get
\[
\big< \phi_{r'} \big| \hat{\varrho} \big| \phi_r \big> = \frac{2}{2N+1} \sum_{m,n=0}^{2N}
\left[
\left(
\sum_{k=0}^{2N} \frac{\exp(ik \phi_m)}{\exp(ik \phi_r) + \exp(ik \phi_{r'})}
  \right)
  \exp \Big\{in (\phi_r- \phi_{r'}) \Big\}\varrho_W(\phi_m,n)
\right]
\]
\be
\label{3.6}
= \frac{2}{2N+1} \sum_{m,n=0}^{2N}
\left[
\left(
\sum_{k=-N}^{N} \frac{\exp(ik \phi_m)}{\exp(ik \phi_r) + \exp(ik \phi_{r'})}
  \right)
  \exp \Big\{in (\phi_r- \phi_{r'}) \Big\}\varrho_W(\phi_m,n)
\right]
\ee
for $ r \geq r'$. 

Hence for $ r \leq r'$ we obtain
\[
\big< \phi_{r'} \big| \hat{\varrho} \big| \phi_r \big> = \big< \phi_{r} \big| \hat{\varrho} \big| \phi_{r'} \big>^*
\]
\be
\label{3.7}
= \frac{2}{2N+1} \sum_{m,n=0}^{2N}
\left[
\left(
\sum_{k=-N}^{N} \frac{\exp(-ik \phi_m)}{\exp(-ik \phi_r) + \exp(-ik \phi_{r'})}
  \right)
  \exp \Big\{-in (\phi_{r'}- \phi_{r}) \Big\}\varrho_W(\phi_m,n).
\right]
\ee
Finally, gathering (\ref{3.6}) and (\ref{3.7}) one concludes that the matrix elements $\big< \phi_{r'} \big| \hat{\varrho} \big| \phi_r \big>,$ $ 0 \leq r, r' \leq 2N$ are given by the formula
\be
\label{3.8}
\big< \phi_{r'} \big| \hat{\varrho} \big| \phi_r \big> = 
\frac{2}{2N+1} \sum_{m,n=0}^{2N}
\left[
\left(
\sum_{k=-N}^{N} \frac{\exp(ik \phi_m)}{\exp(ik \phi_r) + \exp(ik \phi_{r'})}
  \right)
  \exp \Big\{in (\phi_r- \phi_{r'}) \Big\}\varrho_W(\phi_m,n)
\right].
\ee
Employing (\ref{3.8}) we quickly obtain the matrix elements 
$\big< n' \big| \hat{\varrho} \big| n'' \big>,$ $ 0 \leq n', n'' \leq 2N,$ which are of the form
\[
\big< n' \big| \hat{\varrho} \big| n'' \big>= \sum_{r',r''=0}^{2N}
\big<n' \big| \phi_{r'}\big>
\big< \phi_{r'} \big| \hat{\varrho} \big| \phi_{r''} \big>
\big<\phi_{r''} \big| n''\big>
\]
\[
= \frac{2}{(2N+1)^2}
 \sum_{r', r'',m,n=0}^{2N}
\left[
\left(
\sum_{k=-N}^{N} \frac{\exp(ik \phi_m)}{\exp(ik \phi_{r'}) + \exp(ik \phi_{r''})}
  \right) \right.
\]
\be
\label{3.9}
\left. \times
  \exp \Big\{i\big[ (n'-n) \phi_{r'} + (n-n'') \phi_{r''} \big] \Big\}\varrho_W(\phi_m,n)
\right].
\ee
Now we are going to show, how easily one can receive the corresponding formulas of quantum optics in number -- phase variables from the ones presented before. To this end we assume that $N$ tends to infinity, $N \rightarrow \infty$ and substitute
\setcounter{orange}{1}
\renewcommand{\theequation} {\arabic{section}.\arabic{equation}\theorange}
\be
\label{3.10a}
 n \underset{ N \rightarrow \infty}{ \longrightarrow} n \;\;\; , \;\;\;
 \phi_m  \underset{ N \rightarrow \infty}{ \longrightarrow} \phi
\ee
  where $ \phi_0 \leq \phi < \phi_0 + 2 \pi, $
\addtocounter{orange}{1}
\addtocounter{equation}{-1}
\be
\label{3.10b}
\big| n \big> \underset{ N \rightarrow \infty}{ \longrightarrow} \big|n \big> \;\;\; , \;\;\;
\sqrt{\frac{2N+1}{2 \pi} } \big| \phi_m \big> \underset{ N \rightarrow \infty}{ \longrightarrow} 
\big| \phi \big> = \frac{1}{\sqrt{2 \pi}} \sum_{n=0}^{\infty} \exp(i n \phi) \big|n \big>,
\ee
\addtocounter{orange}{1}
\addtocounter{equation}{-1}
\be
\label{3.10c}
\sum_{m=0}^{2N} \frac{2 \pi}{2N+1} f(\phi_m) 
\underset{ N \rightarrow \infty}{ \longrightarrow}
\int_{\phi_0}^{\phi_0 + 2 \pi} f(\phi) d \phi\;\;,\;\;
\sum_{n=0}^{2N}f(n)
\underset{ N \rightarrow \infty}{ \longrightarrow}
\sum_{n=0}^{\infty}f(n),
\ee
\addtocounter{orange}{1}
\addtocounter{equation}{-1}
\be
\label{3.10d}
\frac{2N+1}{2 \pi} \varrho_W(\phi_m,n)
\underset{ N \rightarrow \infty}{ \longrightarrow}
\varrho_W(\phi,n).
\ee
\renewcommand{\theequation} {\arabic{section}.\arabic{equation}}
Using these substitutions we first obtain from (\ref{3.3})
\be
\label{3.11}
\hat{\Omega}(\phi_m,n)
\underset{ N \rightarrow \infty}{ \longrightarrow}
\hat{\Omega}(\phi,n) = \pi \Big( \big|\phi \big> \big<\phi \big|n\big> \big<n \big| +
 \big|n\big> \big<n \big| \phi\big> \big<\phi \big|
  \Big).
\ee
Then, by (\ref{2.18})
\[
\hat{f} = \frac{1}{2N+1} \sum_{m,n=0}^{2N} f(\phi_m,n) \hat{\Omega}(\phi_m,n)
\underset{ N \rightarrow \infty}{ \longrightarrow}
\hat{f} = \frac{1}{2 \pi} \sum_{n=0}^{\infty}
\int_{\phi_0}^{\phi_0+ 2 \pi} f(\phi,n) \hat{\Omega}(\phi,n) d \phi
\]
\be
\label{3.12}
= \frac{1}{2}  \sum_{n=0}^{\infty}
\int_{\phi_0}^{\phi_0+ 2 \pi} f(\phi,n)
 \Big( \big|\phi \big> \big<\phi \big|n\big> \big<n \big| +
 \big|n\big> \big<n \big| \phi\big> \big<\phi \big|
  \Big) d \phi.
\ee
From (\ref{3.5}) with (\ref{3.10b}) and (\ref{3.10d}) one quickly finds that
\be
\label{3.13}
\varrho_W(\phi,n) = \Re \Big\{ \big<n \big|\hat{\varrho} \big| \phi\big>  \big<\phi \big| n\big>\Big\}
\ee
and this is exactly the number -- phase Wigner function introduced in our previous works \cite{PBT14, PB15}.

Then we derive the matrix elements $\big< \phi' \big|\hat{\varrho} \big| \phi''\big>$ from (\ref{3.8})
\[
\big< \phi_{r'} \big|\hat{\varrho} \big| \phi_r\big>
\underset{ N \rightarrow \infty}{ \longrightarrow}
\big< \phi' \big|\hat{\varrho} \big| \phi''\big>
\]
\be
\label{3.14}
= \frac{1}{\pi}  \sum_{n=0}^{\infty} \sum_{k=-\infty}^{\infty} \int_{\phi_0}^{\phi_0+ 2 \pi} d \phi\, 
\varrho_W(\phi,n) \exp \Big\{i n (\phi'' - \phi' ) \Big\}
\frac{\exp(ik \phi)}{\exp(ik \phi') + \exp(ik \phi'')}.
\ee
Finally, Eq. (\ref{3.9}) in the limit $N \rightarrow \infty$ turns into
\[
\big< n' \big|\hat{\varrho} \big| n''\big>
\underset{ N \rightarrow \infty}{ \longrightarrow}
\big< n' \big|\hat{\varrho} \big| n''\big>
\]
\[
= \frac{1}{2\pi^2}  \sum_{n=0}^{\infty} \sum_{k=-\infty}^{\infty} \int_{\phi_0}^{\phi_0+ 2 \pi} d \phi
\int_{\phi_0}^{\phi_0+ 2 \pi} d \phi'
\int_{\phi_0}^{\phi_0+ 2 \pi} d \phi''
\varrho_W(\phi,n) 
\]
\be
\label{3.15}
\times
 \exp \Big\{i \big[ (n'-n)  \phi'  + (n-n'')  \phi''  \big] \Big\}
 \frac{\exp(ik \phi)}{\exp(ik \phi') + \exp(ik \phi'')}.
\ee
Substituting $n' < n''=n' + p, \, p \geq 1$ and introducing $\psi= \phi'' - \phi'$ one gets
\[
\big< n' \big|\hat{\varrho} \big| n' + p\big>= 
 \frac{1}{2\pi^2}  \sum_{n=0}^{\infty} \sum_{k=-\infty}^{\infty} \int_{\phi_0}^{\phi_0+ 2 \pi} d \phi
\int_{\phi_0}^{\phi_0+ 2 \pi} d \phi''
\int_{\phi''-\phi_0-2 \pi}^{\phi''-\phi_0} d \psi \,
\varrho_W(\phi,n) 
\]
\[
\times
\exp(ik \phi) \exp \Big\{-i (p+k)  \phi''   \Big\}
\frac{\exp \Big\{i(n-n') \psi \Big\}}{1+ \exp(-ik \psi)}
\]
\[
= 2   \sum_{n=0}^{\infty} \sum_{k=-\infty}^{\infty} \int_{\phi_0}^{\phi_0+ 2 \pi} d \phi 
\left\{
\varrho_W(\phi,n) \exp(ik \phi) \delta_{-p,k} \sum_{l=0}^{\infty} (-1)^l \delta_{n-n',kl}
\right\}
\]
\be
\label{3.16}
= 2  \int_{\phi_0}^{\phi_0+ 2 \pi} d \phi \,
\exp(-i p \phi) \sum_{l=0}^{\left[ \frac{n'}{p}\right]} (-1)^l  \varrho_W(\phi,n'-pl).
\ee
Putting $n' \rightarrow n, $ $p \rightarrow m$ and $\phi_0 = - \pi$ we recover the formulas (2.13) (for $1 \leq m \leq n$)
and (2.12a) (for $m \geq n+1$) of Ref. \cite{PB15}.

Finally, for $n''=n'$ the formula (\ref{3.15}) reads
\[
\big< n' \big|\hat{\varrho} \big| n' \big>= 
\frac{1}{2\pi^2}  \sum_{n=0}^{\infty} \sum_{k=-\infty}^{\infty} \int_{\phi_0}^{\phi_0+ 2 \pi} d \phi
\int_{\phi_0}^{\phi_0+ 2 \pi} d \phi''
\int_{\phi''-\phi_0-2 \pi}^{\phi''-\phi_0} d \psi \,
\varrho_W(\phi,n) 
\]
\[
\times
\exp(ik \phi) \exp(-ik \phi'')
\frac{\exp \Big\{i(n-n') \psi \Big\}}{1+ \exp(-ik \psi)}
\]
\[
= \frac{1}{\pi} \sum_{n=0}^{\infty} \sum_{k=-\infty}^{\infty} \int_{\phi_0}^{\phi_0+ 2 \pi} d \phi
\int_{\phi_0}^{\phi_0+ 2 \pi} d \psi \,
\varrho_W(\phi,n)  \frac{\exp \Big\{i(n-n') \psi \Big\}}{1+ \exp(-ik \psi)} \delta_{k0}
\]
\be
\label{3.17}
= \int_{\phi_0}^{\phi_0+ 2 \pi} d \phi \, \varrho_W(\phi,n')
\ee
so for $\phi_0=- \pi$ we recover Eq. (2.12c) of \cite{PB15}.

The natural question arises, what is a relationship between our Wigner function on the grid $\Gamma^{(2N+1)}$ and the Wigner function introduced by W. K. Wootters in his distinguished paper \cite{W87} and then by O. Cohendet {\it et al} \cite{CCSS88}, J. A. Vaccaro and D. T. Pegg \cite{VP90} and U. Leonhardt \cite{L95, L96}. We consider this problem in the next section.
%%%%%%%%%%%%%%%%%%%%%%%%%%%%%%%%%%%%%%%%%%%%%%%%%%%%%%%%
%%%%%%%%%%%%%%%%%%%%%%%%%%%%%%%%%%%%%%%%%%%%%%%%%%%%%%%%
\section{The relation between $\varrho_W(\phi_m,n) $ and other discrete Wigner functions in the odd -- dimensional case}
\label{sec4}
\setcounter{equation}{0}

First we define the {\it line $L(n_1,n_2,n_3)$ in $\Gamma^{(2N+1)}$}, $0 \leq n_1, n_2, n_3 \leq 2N$ as a set of points $(\phi_m,n) \in \Gamma^{(2N+1)}$ given by 
\be
\label{4.1}
L(n_1,n_2,n_3)= \Big\{ (\phi_m,n) \in \Gamma^{(2N+1)}: n_1 m +n_2 n = n_3 \; {\rm mod} (2N+1)\Big\}.
\ee
We say that two lines $L(n_1,n_2,n_3)$ and $L(n_1',n_2',n_3')$ are parallel iff $n_1'=n_1$ and $n_2=n_2'$. One quickly finds that using the concept of line we can rewrite Eqs. (\ref{2.28a}) -- (\ref{2.28c}) as
\setcounter{orange}{1}
\renewcommand{\theequation} {\arabic{section}.\arabic{equation}\theorange}
\be
\label{4.2a}
\frac{1}{2N+1} \, \sum_{(\phi_m,n) \in L(1,0,n_3)} \hat{\Omega}[{\cal K}] (\phi_m,n)= \big| \phi_{n_3} \big> \big< \phi_{n_3} \big|,
\ee
\addtocounter{orange}{1}
\addtocounter{equation}{-1}
\be
\label{4.2b}
\frac{1}{2N+1}\, \sum_{(\phi_m,n) \in L(0,1,n_3)} \hat{\Omega}[{\cal K}] (\phi_m,n)= \big| n_3 \big> \big< n_3 \big|,
\ee
\addtocounter{orange}{1}
\addtocounter{equation}{-1}
\be
\label{4.2c}
\frac{1}{2N+1}  \sum_{n_3=0}^{2N} \,
\sum_{(\phi_m,n) \in L(1,0,n_3)} 
\hat{\Omega}[{\cal K}] (\phi_m,n)= 
\frac{1}{2N+1}  \sum_{n_3=0}^{2N} \,
\sum_{(\phi_m,n) \in L(0,1,n_3)} 
\hat{\Omega}[{\cal K}] (\phi_m,n)=
\hat{\bf 1}.
\ee
\renewcommand{\theequation} {\arabic{section}.\arabic{equation}}
It is obvious that in the same way, mutatis mutandi, one can rewrite Eqs. (\ref{2.41}),  (\ref{2.42}) and (\ref{2.40}).

Motivated by Eqs. (\ref{4.2a}) -- (\ref{4.2c})  and employing the pioneering results of W. K. Woothers \cite{W87} and also many others \cite{CCSS88, VP90, L95, L96, MLI01, HTHH02, MPS02} we are looking for such a kernel ${\mathcal K}$ that for any line $L(n_1,n_2,n_3)$ the operator
\be
\label{4.3}
\hat{P}_{L(n_1,n_2,n_3)} = \frac{1}{2N+1} \, \sum_{(\phi_m,n) \in L(n_1,n_2,n_3)} \hat{\Omega}[{\cal K}] (\phi_m,n)
\ee
is a projective operator and 
\be
\label{4.4}
\sum_{n_3=0}^{2N} \hat{P}_{L(n_1,n_2,n_3)}= \hat{\bf 1} \;\; \;\forall \; \; 0 \leq n_1, n_2 \leq 2N.
\ee
In our further considerations we follow \cite{MPS02}. Eq. (\ref{4.3}) can be rewritten as
\[
\hat{P}_{L(n_1,n_2,n_3)} = \frac{1}{2N+1} \, \sum_{m,n=0 }^{2N} \hat{\Omega}[{\cal K}] (\phi_m,n) \delta_{n_1 m + n_2 n - n_3, 0 \; {\rm mod} (2N+1)}
\]
\be
\label{4.5}
= \frac{1}{(2N+1)^2} \, \sum_{r,m,n=0 }^{2N} \hat{\Omega}[{\cal K}] (\phi_m,n)
\exp \left\{ i \frac{2 \pi r}{2N+1}(n_1 m + n_2 n - n_3)\right\}.
\ee
Inserting (\ref{2.17}) into (\ref{4.5}) one has
\[
\hat{P}_{L(n_1,n_2,n_3)} = \frac{1}{(2N+1)^3} \, \sum_{k,l,r,m,n=0 }^{2N} {\cal K}(k,l)
\exp \left\{ i \frac{2 \pi }{2N+1}\big[(n_1 r -k)m + (n_2 r -l)n \big] \right\}
\]
\be
\label{4.6}
 \times \exp \left(-i \frac{2 \pi }{2N+1} r n_3 \right) \hat{D}_0 (k,l),
\ee
where (see (\ref{2.11}))
\be
\label{4.7}
\hat{D}_0 (k,l):= \exp(-ik \phi_0)\hat{D}(k,l), \;\; k,l \in {\mathbb Z}.
\ee
Summation with respect to $m$ and $n$ in (\ref{4.6}) leads to
\[
\hat{P}_{L(n_1,n_2,n_3)} = \frac{1}{2N+1} \, \sum_{k,l,r=0 }^{2N} {\cal K}(k,l)
\delta_{n_1 r -k,0 \; {\rm mod} (2N+1)}
\delta_{n_2 r -l,0 \; {\rm mod} (2N+1)}
\]
\[
 \times \exp \left(-i \frac{2 \pi }{2N+1} r n_3 \right) \hat{D}_0 (k,l)
\]
\[
=
  \frac{1}{2N+1} \, \sum_{r=0 }^{2N} {\cal K} \big( n_1 r\; {\rm mod} (2N+1),  n_2 r\; {\rm mod} (2N+1) \big)
   \exp \left(-i \frac{2 \pi }{2N+1} r n_3 \right) 
\]
\be
\label{4.8}
\times \hat{D}_0 \big( n_1 r\; {\rm mod} (2N+1),  n_2 r\; {\rm mod} (2N+1) \big).
\ee
We put now
\bea
\label{4.9}
n_1 r\; {\rm mod} (2N+1)= n_1 r - \alpha_r (2N+1),&  \nonumber \\
n_1 r\; {\rm mod} (2N+1)= n_2 r - \beta_r (2N+1), & \;\;\;\alpha_r, \beta_r \in {\mathcal N}.
\eea
Then Eq. (\ref{4.8}) takes the form
\[
\hat{P}_{L(n_1,n_2,n_3)} = \frac{1}{2N+1} \,  \sum_{r=0 }^{2N} {\cal K} \big(n_1 r - \alpha_r (2N+1),  n_2 r - \beta_r (2N+1)  \big)
\]
\[
\times
 \exp \left(-i \frac{2 \pi }{2N+1} r n_3 \right) 
  \hat{D}_0  \big(n_1 r - \alpha_r (2N+1),  n_2 r - \beta_r (2N+1)  \big)
\]
\[
= \frac{1}{2N+1} \,  \sum_{r=0 }^{2N} {\cal K} \big(n_1 r - \alpha_r (2N+1),  n_2 r - \beta_r (2N+1)  \big)
 \exp \left(-i \frac{2 \pi }{2N+1} r n_3 \right) 
\]
\be
\label{4.10}
\times (-1)^{\alpha_r \beta_r (2N+1)- \beta_r n_1 r - \alpha_r n_2 r } \hat{D}_0 (n_1 r, n_2 r).
\ee
Since
\be
\label{4.11}
\hat{D}_0 (n_1 r, n_2 r)= \Big( \hat{D}_0 (n_1 , n_2 )\Big)^r,
\ee
Eq. (\ref{4.10}) turns into
\[
\hat{P}_{L(n_1,n_2,n_3)} 
= \frac{1}{2N+1} \,  \sum_{r=0 }^{2N} {\cal K} \big(n_1 r - \alpha_r (2N+1),  n_2 r - \beta_r (2N+1)  \big)
 \exp \left(-i \frac{2 \pi }{2N+1} r n_3 \right) 
\]
\be
\label{4.12}
\times (-1)^{\alpha_r \beta_r (2N+1)- \beta_r n_1 r - \alpha_r n_2 r } \Big(\hat{D}_0 (n_1 , n_2 ) \Big)^r.
\ee
The operator $\hat{D}_0(n_1, n_2)$ is unitary so it has the following spectral decomposition
\be
\label{4.13}
\hat{D}_0(n_1, n_2)= \sum_{j=0}^{2N} \exp(i u_j) \big| u_j \big> \big< u_j \big|, 
\ee
\[
u_j \in {\mathbb R}, \;
\big< u_j  \big| u_j' \big> = \delta_{jj'},  \; j, j' = 0,1, \ldots, 2N. 
\]
As
\[
\Big( \hat{D}_0(n_1, n_2) \Big)^{2N+1}= \hat{D}_0 \big(n_1(2N+1), n_2(2N+1) \big) 
\]
\be
\label{4.14}
=(-1)^{n_1 n_2 (2N+1)} \hat{\bf 1}=  (-1)^{n_1 n_2 (2N+1)}  \sum_{j=0}^{2N} \big|  u_j  \big> \big< u_j \big|
\ee
so substituting (\ref{4.13}) into (\ref{4.14})  one gets that
\be
\label{4.15}
u_j = \frac{2 \pi}{2N+1} \sigma_j + \pi n_1 n_2, \;\;\; \sigma_j \in {\mathbb Z}.
\ee
Inserting (\ref{4.13}) with (\ref{4.15}) into (\ref{4.12}) we have
\[
\hat{P}_{L(n_1,n_2,n_3)} 
= \frac{1}{2N+1} \,  \sum_{j,r=0 }^{2N} {\cal K} \big(n_1 r - \alpha_r (2N+1),  n_2 r - \beta_r (2N+1)  \big)
\]
\[
\times
 (-1)^{\alpha_r \beta_r (2N+1)- \beta_r n_1 r - \alpha_r n_2 r+ n_1 n_2 r }  \exp \left(i \frac{2 \pi r }{2N+1} (\sigma_j -  n_3) \right) \big|  u_j  \big> \big< u_j \big| 
\]
\[
= \frac{1}{2N+1} \,  \sum_{j,r=0 }^{2N} {\cal K} \big(n_1 r - \alpha_r (2N+1),  n_2 r - \beta_r (2N+1)  \big)
\]
\be
\label{4.16}
\times (-1)^{[n_1 r - \alpha_r (2N+1)][n_2 r - \beta_r (2N+1)]}
 \exp \left(i \frac{2 \pi r }{2N+1} (\sigma_j -  n_3) \right) \big|  u_j  \big> \big< u_j \big|.
\ee
From (\ref{4.16}) one easily concludes that if the kernel ${\cal K}$ is such that
\[
 {\cal K} \big(n_1 r - \alpha_r (2N+1),  n_2 r - \beta_r (2N+1)  \big)= 
 (-1)^{[n_1 r - \alpha_r (2N+1)][n_2 r - \beta_r (2N+1)]}
\]
then
\[
\hat{P}_{L(n_1,n_2,n_3)} 
= \frac{1}{2N+1} \,  \sum_{j,r=0 }^{2N} 
 \exp \left(i \frac{2 \pi r }{2N+1} (\sigma_j -  n_3) \right) \big|  u_j  \big> \big< u_j \big|
\]
\be
\label{4.17}
=\sum_{j=0 }^{2N}  \delta_{\sigma_j -  n_3, 0 \, {\rm mod}(2N+1)} \big|  u_j  \big> \big< u_j \big|
\ee
which means that $\hat{P}_{L(n_1,n_2,n_3)} $ is a projective operator. Moreover
\be
\label{4.18}
\sum_{n_3=0 }^{2N} \hat{P}_{L(n_1,n_2,n_3)} = \sum_{j=0 }^{2N}  \big|  u_j  \big> \big< u_j \big|= \hat{\bf 1}.
\ee
Gathering our results we find that for the kernel
\be
\label{4.19}
{\mathcal K}(k,l)= (-1)^{kl}, \;\;\; 0 \leq k,l \leq 2N
\ee
the operator (\ref{4.3}) is projective and the condition (\ref{4.4}) is also satisfied.  One quickly finds that the function 
(\ref{4.19})  fulfills also the conditions (\ref{2.20}), (\ref{2.22a}), (\ref{2.25}) and (\ref{2.26}) imposed on any kernel, and, moreover, it fulfills also the relation (\ref{2.31}) and this fact implies that for the kernel  ${\mathcal K}$ given by (\ref{4.19}) Eq. (\ref{2.30}) with $s+1= 2N+1$ holds true.

Inserting the kernel (\ref{4.19}) into (\ref{2.17}) with $s+1= 2N+1$ and employing (\ref{2.14}), after performing some simple manipulations we get
\[
\hat{\Omega}[(-1)^{kl}](\phi_m,n) = \sum_{p=0 }^{2N}
\exp \left(-i \frac{4 \pi  }{2N+1} pn \right) \big| \phi_{m+p} \big> \big< \phi_{m-p} \big|
\]
\be
\label{4.20}
= \sum_{p=-N }^{N} \exp \left(-i \frac{4 \pi  }{2N+1} pn \right) \big| \phi_{m+p} \big> \big< \phi_{m-p} \big|.
\ee
The matrix representation of $\hat{\Omega}[(-1)^{kl}](\phi_m,n) $ reads
\be
\label{4.21}
\big< n' \big| \hat{\Omega}[(-1)^{kl}](\phi_m,n) \big| n'' \big>= 
\delta_{2n, n' + n'' \, {\rm mod}(2N+1)} \exp \Big\{ i (n' - n'')\phi_m \Big\}.
\ee
Finally, substituting (4.20) into (2.37) one finds that the Wigner function $\varrho_W\big[(-1)^{kl}\big](\phi_m,n)$ equals
\[
\varrho_W \big[(-1)^{kl}\big](\phi_m,n)= 
\frac{1}{2N+1}
\sum_{p=0 }^{2N}
\exp \left(-i \frac{4 \pi  }{2N+1} pn \right) \big< \phi_{m-p} \big|\hat{\varrho} \big|\phi_{m+p} \big>
\]
\be
\label{4.22}
=\frac{1}{2N+1}
\sum_{p=-N }^{N}
\exp \left(-i \frac{4 \pi  }{2N+1} pn \right) \big< \phi_{m-p} \big|\hat{\varrho} \big|\phi_{m+p} \big>.
\ee
Thus we recover the analogous results given in Refs. \cite{W87, CCSS88, VP90, L95, L96, MLI01}.

From (\ref{4.22}) or by using directly (\ref{4.21}) in (\ref{2.37}) one can rewrite the Wigner function in the following form
\[
\varrho_W \big[(-1)^{kl}\big](\phi_m,n)=  \frac{1}{(2N+1)^2} \sum_{n', n''=0 }^{2N} \left(
 \sum_{p=-N }^{N}
\exp \left(i \frac{2 \pi p  }{2N+1} (n' + n'' -2n) \right)  \right.
\]
\[
\left.
\times
\exp \Big\{i (n''-n') \phi_m \Big\} \big< n' \big|\hat{\varrho} \big| n''\big>
\right)
\]
\be
\label{4.23}
 =  \frac{1}{2N+1} \sum_{r=-n }^{n} \exp (i 2 r \phi_m)
\big< n -r  \big|\hat{\varrho} \big| n +r\big>.
\ee
Using the rules (\ref{3.10a}) -- (\ref{3.10d}) and denoting
\[
\frac{2N+1}{2 \pi} \varrho_W \big[(-1)^{kl} \big](\phi_m,n)  \underset{ N \rightarrow \infty}{ \longrightarrow} \widetilde{W}(\phi, n)
\] 
we calculate the limit $N \rightarrow \infty$ of (\ref{4.23}) as
\[
 \widetilde{W}(\phi, n)= \frac{1}{(2 \pi)^2} \sum_{n', n''=0}^{\infty} \Big(\int_{- \pi}^{\pi} \exp \Big\{ i(n' +n'' - 2n) \Theta\Big\} d \Theta \Big) \exp  \Big\{i (n'' - n') \phi \Big\} \big< n' \big| \hat{\varrho} \big| n'' \big>
\]
\[
= \frac{1}{2 \pi} \sum_{n'=0}^{2n} \exp  \Big\{i2 (n - n') \phi \Big\} \big< n' \big| \hat{\varrho} \big| 2n-n' \big>
\]
\be
\label{4.24}
= \frac{1}{2 \pi} \sum_{r=-n}^{n} \exp  (i2 r \phi ) \big< n - r \big| \hat{\varrho} \big| n+r \big>.
\ee
Analogously from (\ref{4.22}) one has
\be
\label{4.25}
 \widetilde{W}(\phi, n)= \int_{- \pi}^{\pi} \exp  (-i2  \Theta )  \big< \phi - \Theta \big| \hat{\varrho} \big| \phi + \Theta \big> d \Theta,
\ee
where $ \big| \phi \pm \Theta \big> \equiv \big| \phi \pm \Theta \, {\rm mod} \,2 \pi\big> $.

So it seems that the `natural' number -- phase Wigner function should be the function $\widetilde{W}(\phi, n)$ defined by (\ref{4.24}) or, equivalently, by (\ref{4.25}). However, one quickly realises that the function $ \widetilde{W}(\phi, n),$ in general, does not fulfill the marginal distribution condition stating that $\sum_{n=0}^{\infty}\widetilde{W}(\phi, n)= \big<\phi \big| \hat{\varrho} \big| \phi \big>.$ Indeed, take
\be
\label{4.26}
\hat{\varrho}= \frac{1}{2} \left( \big|0 \big> + \big| 1 \big> \right) \left(\big< 0 \big| + \big< 1 \big| \right).
\ee
Substituting (\ref{4.26}) into (\ref{4.24}) we obtain
\be
\label{4.27}
\widetilde{W}(\phi, n)=  \frac{1}{4 \pi} \left( \delta_{n0} + \delta_{n1}\right).
\ee
Hence
\be
\label{4.28}
\sum_{n=0}^{\infty}\widetilde{W}(\phi, n) =  \frac{1}{2 \pi} \neq 
\Big< \phi \Big|  \frac{1}{2} \left( \big|0 \big> + \big| 1 \big> \right) \left(\big< 0 \big| + \big< 1 \big| \right) \Big| \phi \Big>= 
\frac{1}{2 \pi} + \frac{1}{2 \pi} \cos \phi \; ({\rm in \;\; general}).
\ee
This proves that the respective marginal distribution condition is not satisfied and, consequently, the function  $\widetilde{W}(\phi, n) $ obtained as the limit 
$\widetilde{W}(\phi, n)  = \lim_{N \rightarrow \infty} \frac{2N+1}{2 \pi} \varrho_W \big[(-1)^{kl}\big](\phi_m,n) $  cannot represent a number -- phase Wigner function.  Way out of this problem proposed by A. Luk\v s and V. Pe\v rinov\'a \cite{LP93, PLP98} consists in permitting $n$ and $r$ be half -- integer numbers, $n,r= 0, \frac{1}{2}, 1 , \frac{3}{2},2 , \ldots$ so that $r$ in the sum (\ref{4.24}) is half -- odd if $n$ is half -- odd. It is clear that in the  Luk\v s --  Pe\v rinov\'a formalism the unphysical half -- odd photon number ghosts are introduced. To avoid this need J. Vaccaro \cite{V95} considers the limit $N \rightarrow \infty$ by adding some term to $\varrho_W \big[(-1)^{kl}\big](\phi_m,n). $ As we have demonstrated in Sec. \ref{sec3} by taking the kernel (\ref{3.2}), which leads to the discrete Wigner function (\ref{3.5}) and then by performing the limit $N \rightarrow \infty$ one gets the number -- phase Wigner distribution (\ref{3.13}) without need of any ghosts or additional terms.

At the end of this section we find the relation between $\varrho_W(\phi_m,n)$ given by (\ref{3.5}) and $\varrho_W \big[(-1)^{kl} \big](\phi_m,n)$. Inserting (\ref{2.48}) with ${\mathcal K}= (-1)^{kl}$ and $s=2N$ into (\ref{3.5}), and then employing (\ref{4.20}) one gets
\[
\varrho_W(\phi_m,n)= \sum_{m', n'=0}^{2N} \Re \Big\{ \big< n \big| \hat{\Omega}\big[(-1)^{kl}\big](\phi_{m'},n') \big|\phi_m \big> \big< \phi_m \big| n \big>\Big\}\varrho_W \big[(-1)^{kl} \big](\phi_{m'},n')
\]
\[
= \sum_{m', n'=0}^{2N} \Re
\left\{   
\sum_{p=-N}^N \exp \left\{-i \frac{4 \pi}{2N+1}p n' \right\} \big< n \big| \phi_{m' + p} \big> \big< \phi_{m'-p} \big| \phi_m \big> \big< \phi_m \big| n \big>
\right\} \varrho_W \big[(-1)^{kl} \big](\phi_{m'},n')
\]
\be
\label{4.29}
= \frac{1}{2N+1} \sum_{m', n'=0}^{2N} 
\cos \left[ \frac{4 \pi}{2N+1} (m-m')(n-n')\right] \varrho_W \big[(-1)^{kl} \big](\phi_{m'},n').
\ee
Multiplying both sides of (\ref{4.29})  by $\frac{2N+1}{2 \pi}$ and taking the limit $N \rightarrow \infty$ we reach a {\bf false} formula
\be
\label{4.30}
\varrho_W(\phi, n)= \frac{1}{2 \pi} \sum_{n'=0}^{\infty} \int_{\phi_0}^{\phi_0 + 2 \pi}
\cos \big[ 2 (\phi - \phi')(n - n')\big] \widetilde{W}(\phi', n') d \phi' \;\;{\rm (false)}
\ee
where $\varrho_W(\phi, n)$ is defined by (\ref{3.5}) and $\widetilde{W}(\phi, n)$ by (\ref{4.24}). However, some straightforward calculations show that if we let $n'$ be half -- integer numbers,  $n'= 0, \frac{1}{2}, 1 , \frac{3}{2},2 , \ldots$ then the following equation holds true
\be
\label{4.31}
\varrho_W(\phi, n)= \frac{1}{2 \pi} \sum_{n'=0, \frac{1}{2}, 1 , \frac{3}{2},2 , \ldots}
 \int_{\phi_0}^{\phi_0 + 2 \pi}
\cos \big[ 2 (\phi - \phi')(n - n')\big] W(\phi', n') d \phi'
\ee
(note that still $n= 1,2,3, \ldots$) with
\[
W(\phi',n') = \frac{1}{2 \pi} \sum_{r=-n'}^{n'} {\displaystyle '} \exp(i 2 r \phi') \big< n'-r \big| \hat{\varrho} \big| n'+r \big>
\]
\be
\label{4.32}
=
 \frac{1}{2 \pi} \int_{-\pi}^{\pi} \exp(-i 2 n' \Theta)\big< \phi' - \Theta \big| \hat{\varrho} \big|\phi' + \Theta \big>
 d \Theta,\;\;\; n'=  0, \frac{1}{2}, 1 , \frac{3}{2},2 , \ldots
\ee
being the Luk\v s --  Pe\v rinov\'a number -- phase Wigner function \cite{LP93, PLP98}, where the symbol $\sum {\displaystyle '}$ means that the summation step is $\frac{1}{2}.$
%%%%%%%%%%%%%%%%%%%%%%%%%%%%%%%%%%%%%%%%%%%%%%%%%%%%%%%%
%%%%%%%%%%%%%%%%%%%%%%%%%%%%%%%%%%%%%%%%%%%%%%%%%%%%%%%%
\section{Even -- dimensional Hilbert spaces}
\label{sec5}
\setcounter{equation}{0}
Here we deal with the case when $s=2N-1$ i.e. $\dim {\mathcal H}^{(s+1)}=2N, \;\; N=1,2,\ldots$ We are going to define the kernel analogous to that given by (\ref{3.2}) but now one meets a problem, since the function $f= \cos \left(\frac{\pi kl}{s+1} \right)$ with $s+1=2N, \; 0 \leq k,l \leq 2N-1$ vanishes for $kl=N(2r+1), \; r=0,1,\ldots$ Thus we need to modify slightly the form of the kernel (\ref{3.2}). Our proposition is the following
\be
\label{5.1}
{\mathcal K}(k,l)= \frac{\cos \left( \frac{\pi kl}{2N}+ \epsilon_N\right)}{\cos \epsilon_N}=
\cos \left( \frac{\pi kl}{2N}\right) - \tan  \epsilon_N \cdot \sin \left( \frac{\pi kl}{2N}\right) ,
\ee
where $\epsilon_N \in {\mathbb R}$ depends on $N$ and is chosen in such a way that $\cos \left( \frac{\pi kl}{2N}+ \epsilon_N\right) \neq 0$ for all $0 \leq k,l \leq 2N-1$ and $\cos \epsilon_N \neq 0.$ For example one can substitute $\epsilon_N = \frac{1}{(2N)^r},\; r=0,1, \ldots$ We quickly show that the function (\ref{5.1}) fulfills all the conditions (\ref{2.20}),  (\ref{2.22a}), (\ref{2.25}) and (\ref{2.26}). Therefore it can be considered as the kernel defining the respective phase -- point operators $\hat{\Omega}[{\mathcal K}].$
Further on we assume that the kernel is represented by the expression  (\ref{5.1}) and we will omit the symbol $[{\mathcal K}]$ simply writing $\hat{\Omega}, \,\hat{\varrho}_W$ instead of $\hat{\Omega}[{\mathcal K}],\, \hat{\varrho}_W[{\mathcal K}].$

Inserting (\ref{5.1}) into (\ref{2.17}) and using (\ref{2.14}) one gets
\[
\hat{\Omega}(\phi_m,n)= N \big(\big| \phi_m \big> \big< \phi_m \big| n \big> \big< n  \big|+ \big| n \big> \big< n \big| \phi_m \big> \big< \phi_m \big| \big)
\]
\be
\label{5.2}
+ i N \tan \epsilon_n \big(\big| \phi_m \big> \big< \phi_m \big| n \big> \big< n  \big|- \big| n \big> \big< n \big| \phi_m \big> \big< \phi_m \big| \big)
\ee
(compare with (\ref{3.3})).

Given (\ref{5.2}) we easily find that for any function $f(\phi_m,n)= f_1(\phi_m) f_2(n)$ the respective operator defined by (\ref{2.18}) is
\[
\hat{f}= \frac{1}{2N}\sum_{m,n=0}^{2N-1} f_1(\phi_m) f_2(n) \hat{\Omega}(\phi_m,n)
\]
\be
\label{5.3}
= \frac{1}{2} \Big( f_1(\hat{\phi}) f_2(\hat{n}) + f_2(\hat{n})  f_1(\hat{\phi})\Big) + \frac{i}{2} \tan \epsilon_n [f_1(\hat{\phi}),f_2(\hat{n})],
\ee
where $[f_1(\hat{\phi}),f_2(\hat{n})]$ stands for the commutator. So taking $|\epsilon_N|$ sufficiently small (but $ \neq 0$)  one can consider (\ref{5.3}) as an {\it almost symmetric ordering}.

Then the Wigner function for the kernel (\ref{5.2}) reads
\[
\varrho_W(\phi_m,n) = \frac{1}{2N} {\rm Tr} \big\{\hat{\varrho} \,\hat{\Omega} (\phi_m,n)\big\}=
\Re \left\{ \big<n \big| \hat{\varrho} \big| \phi_m \big> \big< \phi_m \big| n \big> \right\}
- \tan \epsilon_N \Im  \left\{ \big<n \big| \hat{\varrho} \big| \phi_m \big> \big< \phi_m \big| n \big> \right\}
\] 
\be
\label{5.4}
= \frac{1}{\cos \epsilon_N} \Re \left\{ \exp(i \epsilon_N )\big<n \big| \hat{\varrho} \big| \phi_m \big> \big< \phi_m \big| n \big> \right\}
\ee
(compare with (\ref{3.5})).

Employing the general formulas (\ref{2.44}) and (\ref{2.45}) or (\ref{2.46})  we can find a state $\hat{\varrho}$ from the corresponding Wigner function (\ref{5.4}).
\newline
\underline{Example } \hspace{1cm} The qubit

Here we consider  the $2$  -- dimensional  Hilbert space ${\mathcal H}^{(2)}$ so $N=1.$ Straightforward calculations lead to the phase -- point operator (\ref{5.2}) of the form
\[
\hat{\Omega}(\phi_m,n)= \frac{1}{2}
\Big[ \hat{\bf 1}+ (-1)^m
 \big(\big| \phi_0 \big> \big< \phi_0 \big|-   \big| \phi_1 \big> \big< \phi_1 \big| \big)
 + (-1)^n \big(\big| \phi_0 \big> \big< \phi_1 \big| +  \big| \phi_1 \big> \big< \phi_0 \big| \big)
\]
\be
\label{5.5}
+ i (-1)^{m+n} \tan \epsilon_1 \big(\big| \phi_0 \big> \big< \phi_1 \big| -  \big| \phi_1 \big> \big< \phi_0 \big| \big)
\Big]
\ee
or, as the matrix operator in the basis 
$ \big| 0 \big>
 =\left( \begin{array}{c}
 1 \\ 0
 \end{array} \right),
 $
 $ \big| 1 \big>
 =\left( \begin{array}{c}
 0 \\ 1
 \end{array} \right),
 $
 \be
 \label{5.6}
 \hat{\Omega}(\phi_m,n)= \frac{1}{2} \left[ \hat{\bf 1} + (-1)^n \sigma_3 
 + \left(
  \begin{array}{cc}
 \exp(-i \phi_0) & 0  \\ 0&\exp(i \phi_0)
 \end{array} 
 \right) \cdot \Big( (-1)^m \sigma_1 +(-1)^{m+n} \tan \epsilon_1 \cdot \sigma_2\Big)
 \right],
 \ee
 where 
 \[
 \sigma_1=  \left(
  \begin{array}{cc}
0 & 1  \\ 1&0
 \end{array}  \right)\; , \; 
  \sigma_2=  \left(
  \begin{array}{cc}
0 & -i  \\ i&0
 \end{array}  \right)\; , \;
  \sigma_3=  \left(
  \begin{array}{cc}
1 & 0  \\ 0&-1
 \end{array} \right)
 \]
 are the Pauli matrices. 
 Note that with $\epsilon_1= \frac{\pi}{4}$ and $\phi_0=0$ the phase -- point operator (\ref{5.6}) is exactly the one found by  W. K. Wootters \cite{W87} and then by S. Chaturvedi {\it et al} \cite{CEMMMS06}. 
 
 A general form of the density matrix in $2$  -- D case is
 \be
 \label{5.7}
 \hat{\varrho} = \frac{1}{2} \left( \hat{\bf 1} + \vec{a} \cdot \vec {\sigma} \right),
 \ee
 where $\vec{a}=(a_1, a_2, a_3) \in {\mathbb R}^3$, $ a_1^2 + a_2^2 + a_3^2 \leq 1. $
 Substituting (\ref{5.7}) into (\ref{5.4}) and then performing simple algebraic manipulations one finds the Wigner function for the qubit as
 \bea
 \label{5.8}
 \varrho_W(\phi_0,0) & = & 
 \frac{1}{4} \Big[ 
 1 + (a_1 + a_2 \tan \epsilon_1) \cos \phi_0 + (a_2-a_1 \tan \epsilon_1) \sin \phi_0 + a_3
 \Big], \nonumber \\
 \varrho_W(\phi_0,1) & = & 
 \frac{1}{4} \Big[ 
 1 + (a_1 - a_2 \tan \epsilon_1) \cos \phi_0 + (a_2+ a_1 \tan \epsilon_1) \sin \phi_0 - a_3
 \Big], \nonumber \\
 \varrho_W(\phi_1,0) & = & 
 \frac{1}{4} \Big[ 
 1 - (a_1 + a_2 \tan \epsilon_1) \cos \phi_0 - (a_2-a_1 \tan \epsilon_1) \sin \phi_0 + a_3
 \Big], \nonumber \\
 \varrho_W(\phi_1,1) & = & 
 \frac{1}{4} \Big[ 
 1 - (a_1 - a_2 \tan \epsilon_1) \cos \phi_0 - (a_2+a_1 \tan \epsilon_1) \sin \phi_0 - a_3
 \Big].
 \eea
 With $\epsilon_1= \frac{\pi}{4}$ and $\phi_0=0$ we recover the results of Refs. \cite{F87, W87, CEMMMS06}.
 
 Let us return to the general formula (\ref{5.4}). Multiplying both sides of (\ref{5.4}) by $\frac{2N}{2 \pi}$ and taking the limit $N \rightarrow \infty$ under the assumption that $\epsilon_N \underset{N \rightarrow \infty}{\longrightarrow} 0$ ( e.g. $\epsilon_N= \frac{1}{N^p}, \;p=1,2, \ldots$) one gets (\ref{3.13}). An interesting question is what relations are between our Wigner function $\varrho_W(\phi_m,n)$ defined by (\ref{5.4}) and the Wigner function found by U. Leonhardt \cite{L95, L96} (see also \cite{MPS02}) as the one which arises in the tomography of the quantum state in the finite -- dimensional Hilbert space. 
 
 In odd dimensions such a Wigner function is just the function $\varrho_W \big[(-1)^{kl} \big](\phi_m,n)$ analysed in the previous section. In the case of even -- dimensional Hilbert space ${\mathcal H}^{(s+1)}$, $s+1= 2N, \, N=1,2, \ldots$ the problem is more involved and its solution a little bit sophisticated. It is so because when $s+1=2N$ (instead of $s+1=2N+1$) there does not exist, in general, any kernel ${\mathcal K}(k,l)$ satisfying the conditions analogous to (\ref{4.3}) and (\ref{4.4}). (Note that, however such a kernel exists for a qubit $2N=2$ and it reads  ${\mathcal K}(k,l)=(-1)^{kl}$.) A way out of this problem has been proposed by U. Leonhardt and it resembles very much the 
 Luk\v s -- Pe\v rinov\'a idea to define the number -- phase Wigner function \cite{L95, L96, LP93, PLP98}. So in Leonhardt's approach the phase space consists of the points $(\phi_m,n)$ with $m$ and $n$ being half -- integer numbers $m,n= 0, \frac{1}{2}, 1, \frac{3}{2}, \ldots, 2N - \frac{1}{2}. $ Therefore,the dimension of phase space is doubled and is equal to $4N.$ In other words some ghosts are introduced. Then the Wigner function is defined as \cite{L95, L96}
 \[
 W(\phi_m,n)= \frac{1}{4N} \sum_{p=0}^{2N - \frac{1}{2}}{\displaystyle '} \exp\left(-i \frac{4 \pi}{2N}pn \right)
 \big< \phi_{m-p} \big| \hat{\varrho}\big| \phi_{m+p} \big>
 \]
 \[
 =  \frac{1}{4N} \sum_{p=-N}^{N - \frac{1}{2}}{\displaystyle '}  \exp\left(-i \frac{4 \pi}{2N}pn \right)
 \big< \phi_{m-p} \big| \hat{\varrho}\big| \phi_{m+p} \big>
 \]
 \be
 \label{5.9}
 = \frac{1}{4N} {\sum_{r=-n}^{n}}{\displaystyle '} \exp(i 2r \phi_m) \big< n-r \big| \hat{\varrho} \big| n+r \big>.
 \ee
 where, as before in Eq. (\ref{4.32}), the symbol $\sum {\displaystyle '}$ means that the summation step is $\frac{1}{2}.$ Note also that the vectors $\big| \phi_m \big>$ and $\big| n \big>$ vanish if $m$ and $n$ are half -- odd.
 
 Introduce the phase -- point operators \cite{MPS02}
 \[
 \hat{A}(\phi_m,n) = \frac{1}{2}
 \sum_{p=0}^{2N - \frac{1}{2}}{\displaystyle '}  \exp\left(-i \frac{4 \pi}{2N}pn \right)
 \big| \phi_{m+p} \big> \big< \phi_{m-p} \big| 
 \]
 \be
 \label{5.10}
  = \frac{1}{2}
 \sum_{p=-N}^{N - \frac{1}{2}}{\displaystyle '}  \exp\left(-i \frac{4 \pi}{2N}pn \right)
 \big| \phi_{m+p} \big> \big< \phi_{m-p} \big|.
 \ee
 With the use of this operator Eq. (\ref{5.9}) can be rewritten in the form
 \be
 \label{5.11}
 W(\phi_m,n) = \frac{1}{2N} {\rm Tr} \big\{ \hat{\varrho}  \hat{A}(\phi_m,n) \big\}.
 \ee
 Straightforward calculations lead to the inverse formula \cite{MPS02}
 \be
 \label{5.12}
 \hat{\varrho} = \sum_{m,n=0}^{2N - \frac{1}{2}}{\displaystyle '} \; \;W(\phi_m,n)  \hat{A}(\phi_m,n).
 \ee
Inserting (\ref{5.12}) into (\ref{5.4}) after some simple algebraic manipulations one gets a relation between  $W(\phi_m,n) $ and $\varrho_W(\phi_m,n)$
 \[
 \varrho_W(\phi_m,n) = \frac{1}{\cos \epsilon_N } \Re \left\{
 \exp (i \epsilon_N)
 \big< n \big|
  \sum_{m',n'=0}^{2N - \frac{1}{2}}{\displaystyle '} \; \;W(\phi_{m'},n')  \hat{A}(\phi_{m'},n')
  \big| \phi_m \big> \big< \phi_m \big| n \big>
 \right\}
 \]
 \[
 = \frac{1}{\cos \epsilon_N } 
  \sum_{m',n'=0}^{2N - \frac{1}{2}}{\displaystyle '} \; \;W(\phi_{m'},n') 
 \Re \left\{
 \exp (i \epsilon_N)
 \big< n \big|
  \hat{A}(\phi_{m'},n')
  \big| \phi_m \big> \big< \phi_m \big| n \big>
 \right\}
 \]
 \be
 \label{5.13}
 = \frac{1}{4N \cos \epsilon_N } 
 \sum_{m',n'=0}^{2N - \frac{1}{2}}{\displaystyle '}\;\;
 \cos \left[ \frac{4 \pi}{2N} (m - m') (n - n') - \epsilon_N
 \right] W(\phi_{m'},n') 
 \ee
 (compare with (\ref{4.29}) for odd dimensions).
 Formula (\ref{5.13})  shows that from tomography for the Wigner function $W(\phi_m,n)$ one can reconstruct also the Wigner function $ \varrho_W(\phi_m,n) $.
 
 Finally, consider the limit $N \rightarrow \infty.$ From (\ref{5.9}) we find that
 \be
 \label{5.14}
 \frac{4 \pi}{2N} W(\phi_m,n) \underset{N \rightarrow \infty}{\longrightarrow}
 W(\phi_m,n)= \frac{1}{2 \pi}
 \sum_{r=-n}^{n}{\displaystyle '}\;\;
 \exp(i 2r \phi) \big< n-r \big| \hat{\varrho} \big| n+r \big>
 \ee
 so this limit gives exactly the  Luk\v s --  Pe\v rinov\'a number -- phase Wigner function (\ref{4.32}). Now multiplying both sides of (\ref{5.13}) by $ \frac{2N}{2 \pi}$ and taking the  limit $N \rightarrow \infty $, $ \epsilon_N \rightarrow 0 $ one obtains the {\bf false} formula
 \be
 \label{5.15}
 \varrho_W(\phi_m,n)=  \frac{1}{4 \pi}  \sum_{n'=0}^{\infty}{\displaystyle '}\;\;
 \int_{\phi_0}^{\phi_0 + 2 \pi} 
 \cos \left[ 2 (\phi - \phi') (n - n')  \right] 
 W(\phi',n')  d \phi' 
 \;\; {\rm (false)}
 \ee
 (compare with the right formula (\ref{4.31})).
 
 In conclusion we can say that in contrast to the Wigner function for the symmetric ordering where the limit $N \rightarrow \infty $ is almost automatically defined, in the case of the Wigner function $\varrho_W \big[(-1)^{kl} \big](\phi_m,n)$ considered in the preceding section and the Wigner function $W(\phi_{m},n)$ analysed in the previous section the limit $N \rightarrow \infty $ may lead to the false results.
 %%%%%%%%%%%%%%%%%%%%%%%%%%%%%%%%%%%%%%%%%%%%%%%%%%%%%%%%
%%%%%%%%%%%%%%%%%%%%%%%%%%%%%%%%%%%%%%%%%%%%%%%%%%%%%%%%
\section{Summary}
\label{sec6}
In this paper we have developed the general discrete Weyl -- Wigner formalism. Then this formalism has been specified to the case of symmetric ordering of operators (for odd -- dimensional Hilbert spaces) and to the almost symmetric ordering (for even -- dimensional Hilbert spaces). The advantage of this approach is a simple rule of quantisation and also the fact that there is no need  of implementing the `ghosts' when the Wigner function is defined for an even -- dimensional Hilbert space. Moreover, a natural limit when the dimension of the phase -- space grid tends to infinity leads to the number -- phase Wigner function.

The disadvantage of this formalism is  lack of  a direct tomography of the quantum state. However, some indirect tomography can be achieved by the use of Eq. (\ref{4.29}) or (\ref{5.13}) since the tomography of $\varrho_W \big[(-1)^{kl} \big](\phi_m,n)$ or  $W(\phi_{m},n)$ can be done \cite{L96}.

%%%%%%%%%%%%%%%%%%%%%%%%%%%%%%%%%%%%%%%%%%%%%%%%%%%%%%%%
%%%%%%%%%%%%%%%%%%%%%%%%%%%%%%%%%%%%%%%%%%%%%%%%%%%%%%%%

\end{document}